\documentclass[twocolumn,showpacs,preprintnumbers,amsmath,amssymb]{revtex4}
\usepackage{amsmath,amssymb,graphics,epsfig,subfigure}
\usepackage{color}

\begin{document}
\thispagestyle{empty}

\begin{center}

\title{Topology of black hole thermodynamics}

\date{\today}
\author{Shao-Wen Wei\footnote{Corresponding author. E-mail: weishw@lzu.edu.cn},
Yu-Xiao Liu\footnote{E-mail: liuyx@lzu.edu.cn}}

\affiliation{$^{1}$Lanzhou Center for Theoretical Physics, Key Laboratory of Theoretical Physics of Gansu Province, School of Physical Science and Technology, Lanzhou University, Lanzhou 730000, People's Republic of China,\\
 $^{2}$Institute of Theoretical Physics $\&$ Research Center of Gravitation,
Lanzhou University, Lanzhou 730000, People's Republic of China}

\begin{abstract}
A critical point is an important structure in the phase diagram of a thermodynamic system. In this work, we introduce topology to the study of the black hole thermodynamics for the first time by following Duan's topological current $\phi$-mapping theory. Each critical point is endowed with a topological charge. We find that critical points can be divided into two classes, the conventional and the novel. Further study shows that the first-order phase transition can extend from the conventional critical point, while the presence of the novel critical point cannot serve as an indicator of the presence of the first-order phase transition near it. Moreover, the charged anti-de Sitter black hole and the Born-Infeld anti-de Sitter black hole have different topological charges, which indicates they are in different topological classes from the viewpoint of thermodynamics. These give the first promising study on the topology of the black hole thermodynamics. Such approach is also expected to be extended to other black holes, and much more topological information remains to be disclosed.
\end{abstract}


\maketitle
\end{center}

\section{Introduction}

Thermodynamics is one of the most fascinating fields in black hole physics. Apart from the fact that the strong gravitational phenomena can be tested through the observation of the gravitational waves \cite{Abbott}, the black hole thermodynamics remains to be tested. Of particular interest is that the Bekenstein-Hod universal bound on information emission rate was checked in Ref. \cite{Carullo}, which is consistent with black hole thermodynamics. This further confirms that black holes are indeed thermodynamic systems. However, early studies just revealed that there are four laws of black hole mechanics. For example, the first law reads \cite{Bardeen}
\begin{eqnarray}
 dM=\frac{\kappa}{8\pi G} d\mathcal{A}+\sum_{i}Y_i dx^{i},\label{manlaw}
\end{eqnarray}
where $M$, $\kappa$, and $\mathcal{A}$ are, respectively, the mass, surface gravity, and area of the black hole. $Y_i dx^{i}$ is the $i$-th chemical potential term. Toward the thermodynamics, Bekenstein and Hawking made a big step when they regarded the area and surface gravity as the entropy and temperature of the black hole \cite{Hawking,Bekensteina,Bekensteinb},
\begin{equation}
 S=\frac{\mathcal{A}}{4G},\quad T=\frac{\kappa}{2\pi}.
\end{equation}
Naturally, the four laws of black hole mechanics become the laws of  thermodynamics, and (\ref{manlaw}) turns to
\begin{eqnarray}
 dM=TdS+\sum_{i}Y_i dx^{i}.
\end{eqnarray}
It is quite interesting that the phase transition has been discovered in the black hole systems including the Hawking-Page phase transition \cite{Page} and the small-large black hole phase transition \cite{Chamblin}. Especially, recent study indicates that the cosmological constant can be interpreted as the thermodynamics pressure \cite{Kastor}, which leads to
\begin{eqnarray}
 dM=TdS+VdP+\sum_{i}Y_i dx^{i},\label{fila}
\end{eqnarray}
while the mass obtains a new physical meaning, the enthalpy $M\equiv H$ rather than the energy of the system. Subsequently, rich phase transitions and phase structures have been observed. Among them, the small-large black hole phase transition was found to exist in most of the black hole systems. Such phase transition is similar to the liquid-gas phase transition of the van der Waals fluid \cite{Kubiznak}, and the corresponding black hole microstructure and characteristic interaction potential were explored \cite{Weiw,Weiwa2,Weiwa3}. With the decrease of the temperature, the coexistence curve of the small and large black holes extends from a critical point and ends at the origin in the pressure-temperature diagram. Near the critical point, the universal phenomena are observed, which is the same as that of the mean field theory. Although different phase structures possess different patterns, a critical point always emerges, and it plays a key to understand the system. In the following, we shall examine the corresponding topological property corresponding to the critical point.

\section{Thermodynamical function and topology}

Temperature and entropy are two key quantities in black hole thermodynamics. In general, the temperature can be expressed as a function of the entropy $S$, pressure $P$, and other parameters $x^{i}$ for a thermodynamic system
\begin{eqnarray}
 T=T(S, P, x^{i}).\label{tptp}
\end{eqnarray}
Here one can eliminate one parameter in the temperature (\ref{tptp}) by requiring $(\partial_{S}T)_{P,x^{i}}=0$. Then we denote the new thermodynamic function as $\Phi$ with the pressure $P$ being eliminated,
\begin{eqnarray}
 \Phi=\frac{1}{\sin\theta}T(S, x^{i}),\label{phinew}
\end{eqnarray}
where the factor $\frac{1}{\sin\theta}$ is an auxiliary term, which can simplify our study of the topology of the black hole thermodynamics.

Now we introduce a new vector field $\phi=(\phi^S, \phi^\theta)$,
\begin{eqnarray}
 \phi^S=(\partial_{S}\Phi)_{\theta,x^{i}},\quad
 \phi^\theta=(\partial_{\theta}\Phi)_{S,x^{i}}.
\end{eqnarray}
The first advantage of the $\theta$ term is that the direction of the introduced vector $\phi$ is perpendicular to the horizontal lines at $\theta=0$ and $\pi$, which can be treated as two boundaries in the parameter space. Another advantage is that the zero point of $\phi$ is always at $\theta=\pi/2$. It is also easy to check that the critical point exactly locates at the zero point of $\phi$. This is an important property that allows us to introduce the topology to study the black hole critical point.

Here, we would like to give a brief discussion on the conditions determining the critical point. Recently, when treating the cosmological constant as pressure \cite{Kastor}, the phase transition has been extensively studied in various anti-de Sitter (AdS) black hole systems. The small-large black hole phase transition was found to be similar to the liquid-gas phase transition of the van der Waals fluid \cite{Kubiznak}. In particular, in the pressure-temperature plane, this first-order phase transition starts at the origin and ends at a critical point with the increase of the temperature. The critical point is generally determined by
\begin{eqnarray}
 \left(\frac{\partial P}{\partial V}\right)_{T}=0,\quad
 \left(\frac{\partial^2P}{\partial V^2}\right)_{T}=0.\label{cococ}
\end{eqnarray}
However, these conditions are not unique. In Refs. \cite{Weiwbb,Weiwcc}, via starting from the first law of black hole thermodynamics and the free energy, we have pointed out that the small-large black hole phase transition point can be obtained from the Maxwell equal area law in different thermodynamical parameter spaces. As a straightforward derivation, the conditions to determine the critical point are given. For example, if one starts from the first law (\ref{fila}), beside the conditions (\ref{cococ}), there are two more kinds of conditions, i.e.,
\begin{eqnarray}
 \left(\frac{\partial T}{\partial S}\right)_{P, x^{i}}=0,\quad
 \left(\frac{\partial^2T}{\partial S^2}\right)_{P, x^{i}}=0,\label{cots}\\
 \left(\frac{\partial x^{i}}{\partial Y_{i}}\right)_{T, P}=0,\quad
 \left(\frac{\partial^2 x^{i}}{\partial Y_{i}^2}\right)_{T, P}=0.\label{yots}
\end{eqnarray}
As we know, an arbitrary black hole possesses the entropy and temperature, so the conditions (\ref{cots}) are universal, which is not only effective for the charged AdS black holes, but also for the rotating black holes. This is also one of the reasons that we choose this condition to obtain the critical point in this paper. On the other hand, the equivalence between the conditions (\ref{cococ}) and (\ref{cots}) has been confirmed in Ref. \cite{Weiwdd} (see the appendix) by using the mathematical calculation.

Following Duan's $\phi$-mapping topological current theory \cite{Duana,Duanb}, we can construct the topological current as
\begin{eqnarray}
 j^{\mu}=\frac{1}{2\pi}\epsilon^{\mu\nu\rho}\epsilon_{ab}\partial_{\nu}n^{a}\partial_{\rho}n^{b},
 \quad \mu,\nu,\rho=0,1,2,
\end{eqnarray}
where $\partial_{\nu}=\frac{\partial}{\partial x^{\nu}}$ and $x^{\nu}$=($t$, $r$, $\theta$). The normalized vector is defined as $n^a=\frac{\phi^a}{||\phi||}$ ($a$=1, 2) with $\phi^1=\phi^S$ and $\phi^2=\phi^\theta$. It is easy to check that this topological current is conserved
\begin{equation}
 \partial_{\mu}j^{\mu}=0.
\end{equation}
By making use of the Jacobi tensor $\epsilon^{ab}J^{\mu}\left(\frac{\phi}{x}\right)=\epsilon^{\mu\nu\rho}
\partial_{\nu}\phi^a\partial_{\rho}\phi^b$ and the two-dimensional Laplacian Green function $\Delta_{\phi^a}\ln||\phi||=2\pi\delta(\phi)$, this topological current can be further expressed as
\begin{equation}
 j^{\mu}=\delta^{2}(\phi)J^{\mu}\left(\frac{\phi}{x}\right).\label{juu}
\end{equation}
From this expression, it is clear that $j^{\mu}$ is nonzero only at the zero points of $\phi^{a}$, i.e., $\phi^a(x^i)=0$, and we denote its $i$-th solution as $\vec{x}=\vec{z}_{i}$. Then according to the $\delta$-function theory \cite{Schouton}, one can obtain the density of the topological current
\begin{equation}
 j^{0}=\sum_{i=1}^{N}\beta_{i}\eta_{i}\delta^{2}(\vec{x}-\vec{z}_{i}).
\end{equation}
The positive Hopf index $\beta_i$ measures the number of the loops that $\phi^a$ makes in the vector $\phi$ space when $x^{\mu}$ goes around the zero point $z_i$ (we suppose there are $N$ solutions). The Brouwer degree $\eta_{i}=\text{sign}(J^{0}({\phi}/{x})_{z_i})=\pm1$. Finally, the corresponding topological charge at given parameter region $\Sigma$ can be calculated via
\begin{eqnarray}
 Q=\int_{\Sigma}j^{0}d^2x
 =\sum_{i=1}^{N}\beta_{i}\eta_{i}=\sum_{i=1}^{N}w_{i},
\end{eqnarray}
where $w_{i}$ is the winding number for the $i$-th zero point of $\phi$~\footnote{Note that, by treating the winding number as a topological charge, the topology of the light ring in the black hole background has been studied in Refs. \cite{Cunhaa,Cunhab,Wei2019}}.

Obviously, for each critical point, we can endow it with a topological charge, which equals the winding number. For a thermodynamic system, we can calculate its topological charge when one chooses $\Sigma$ as its complete thermodynamic parameter space. Then different thermodynamic systems can be divided into different classes. This shall allow us to examine the topological transition among these different thermodynamic systems. On the other hand, since $\eta_{i}$ can be positive or negative, these critical points have two different topological properties. Here we would like to name them as the conventional one with $\eta_{i}=-1$ (or $w_{i}=-1$) and novel one with $\eta_{i}=1$ (or $w_{i}$=1).

\section{Topology of black hole systems}

The charged AdS black hole is the first black hole system admitting the small-large black hole phase transition \cite{Chamblin,Kubiznak} and a critical point is present in the phase diagram of such system. Here we would like to examine its topological property by calculating the topological charge.

\begin{figure}
\includegraphics[width=7cm]{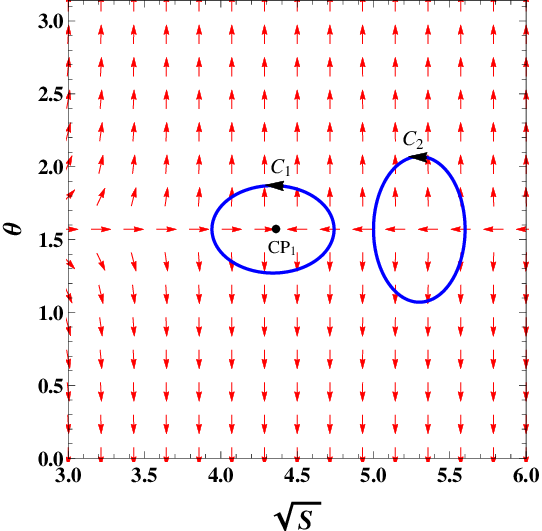}
\caption{The red arrows represent the vector field $n$ on a
portion of the $\sqrt{S}$-$\theta$ plane for the charged AdS black hole with the charge $q$=1. The critical point CP$_1$ located at ($\sqrt{S}$, $\theta$)=($\sqrt{6\pi}q$, $\frac{\pi}{2}$) is marked with a black dot. The blue contours $C_1$ and $C_2$ are two closed loops and $C_1$ encloses the critical point, while $C_2$ does not.}\label{pCharVec}
\end{figure}

There are many approaches to obtain the black hole Hawking temperature. Regarding the inverse of the period of the Euclidean section of the black hole space-time as the Hawking temperature, one obtains
\begin{eqnarray}
 T=\frac{2 P \sqrt{S}}{\sqrt{\pi }}-\frac{\sqrt{\pi } q^2}{4
   S^{\frac{3}{2}}}+\frac{1}{4 \sqrt{\pi } \sqrt{S}},\label{iscop}
\end{eqnarray}
where $q$ and $P$ are the charge and pressure of the black hole system, respectively. The first law holds, i.e., $dM=TdS+\varphi dq+VdP$ with $\varphi$ and $V$ being the electric potential and thermodynamic volume of the black hole system. A little algebra yields the explicit form of the thermodynamic function
\begin{eqnarray}
 \Phi=\frac{1}{\sin\theta}\left(\frac{1}{2 \sqrt{\pi S }}-\frac{\sqrt{\pi }
   q^2}{S^{\frac{3}{2}}}\right).
\end{eqnarray}
The components of the vector field $\phi$ are
\begin{eqnarray}
 \phi^{S}&=&\frac{\csc\theta \left(6\pi q^2-S\right)}{4
   \sqrt{\pi } S^{\frac{5}{2}}},\\
 \phi^{\theta}&=&-\frac{\cot \theta \csc \theta
   \left(S-2 \pi q^2\right)}{2 \sqrt{\pi
   } S^{\frac{3}{2}}}.
\end{eqnarray}
The normalized vector field can be obtained through $n=(\frac{\phi^{S}}{||\phi||}, \frac{\phi^{\theta}}{||\phi||})$, which is exhibited in Fig. \ref{pCharVec}. We can clearly see there is a critical point at $(\sqrt{S},\theta)=(\sqrt{6\pi}q,\frac{\pi}{2})$.

\begin{figure}
\includegraphics[width=7cm]{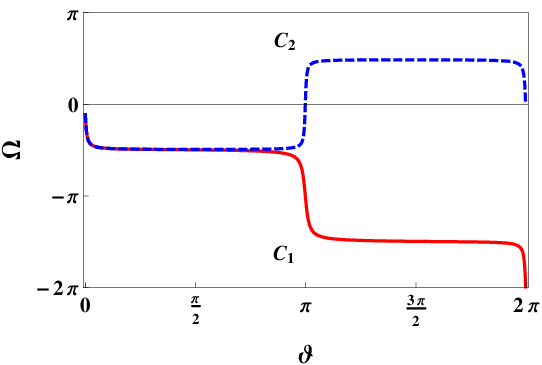}
\caption{$\Omega$ vs $\vartheta$ for contours $C_1$ (red solid curve) and $C_2$ (blue dashed curve).}\label{pOmega}
\end{figure}

From the viewpoint of topology, we know that if the contour encloses the critical point, it will give a nonzero topological charge, otherwise it is zero. In order to calculate the topological charge, we shall construct two contours $C_1$ and $C_2$, which are parametrized by $\vartheta$ $\in$ ($0$, $2\pi$) as
\begin{eqnarray}
\left\{
\begin{aligned}
 r&=a\cos\vartheta+r_0, \\
 \theta&=b\sin\vartheta+\frac{\pi}{2}.
\end{aligned}
\right.\label{pfs}
\end{eqnarray}
We choose $(a, b, r_0)$=(0.4, 0.3, $\sqrt{6\pi}$) for $C_1$, and (0.3, 0.5, 5.3) for $C_2$. We define a new quantity measuring the deflection of the vector field along the given contour
\begin{eqnarray}
 \Omega(\vartheta)=\int_{0}^{\vartheta}\epsilon_{ab}n^{a}\partial_{\vartheta}n^{b}d\vartheta.
\end{eqnarray}
Then the topological charge must be $Q=\frac{1}{2\pi}\Omega(2\pi)$. For the contours $C_1$ and $C_2$, we list $\Omega(\vartheta)$ in Fig. \ref{pOmega}. Considering that the contour $C_2$ does not enclose the critical point, so one must have $Q=0$. From the figure, we clearly see that with the increase of $\vartheta$, $\Omega$ first decreases, then increases, and finally vanishes at $\vartheta=2\pi$. Thus, we get $Q=\frac{1}{2\pi}\Omega(2\pi)=0$ as expected, while for $C_1$, $\Omega$ gradually decreases and approaches $-2\pi$ at $\vartheta=2\pi$. Therefore the topological charge $Q_{\text{CP}_1}=-1$. According to our classification, this critical point is a conventional one. Actually, near this point, there exists a stable small-large black hole phase transition of first order. Moreover, since there exists only one critical point, we have the topological charge
\begin{eqnarray}
 Q=-1,
\end{eqnarray}
for the charged AdS black hole system.

\begin{figure}
\includegraphics[width=7cm]{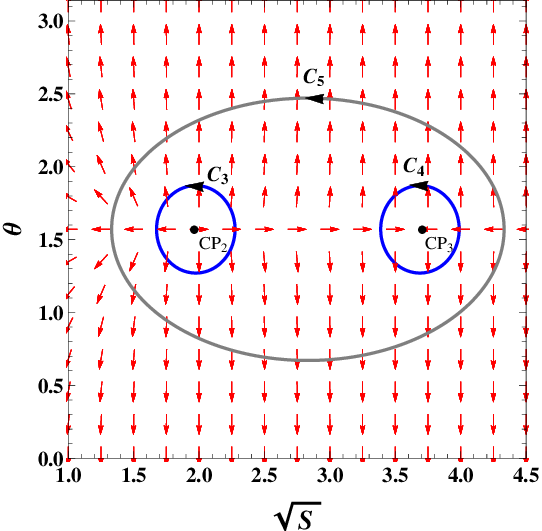}
\caption{The red arrows represent the vector field $n$ on a
portion of the $\sqrt{S}$-$\theta$ plane for the charged BI-AdS black hole with the charge $q$=1 and $b$=0.4. The critical points CP$_2$ and CP$_3$ located at ($\sqrt{S}$, $\theta$)=(1.97, $\frac{\pi}{2}$) and (3.69, $\frac{\pi}{2}$) are marked with black dots, and they are enclosed with the blue contour $C_3$ and $C_4$, respectively, while the gray contour $C_5$ encloses both the critical points.}\label{pBIVec}
\end{figure}

\begin{figure}
\includegraphics[width=7cm]{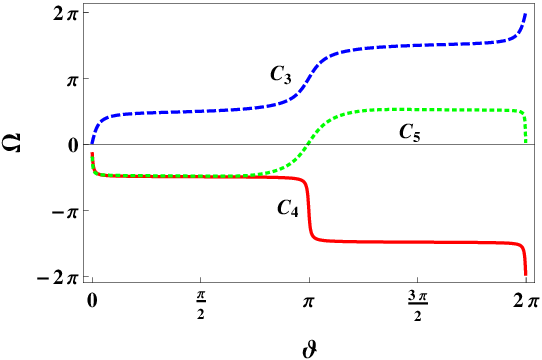}
\caption{$\Omega$ vs $\vartheta$ for the contours $C_3$ (solid curve), $C_4$ (dashed curve), and $C_5$ ( dot dashed curve) for the charged BI-AdS black hole.}\label{pBICh}
\end{figure}

Now let us turn to the charged Born-Infeld (BI) AdS black hole \cite{Fernando}, where besides the conventional critical point, a novel one shall emerge. After treating the cosmological constant as the pressure, the Hawking temperature takes the following form \cite{Gunasekaranad}
\begin{eqnarray}
 T=\frac{1}{4\sqrt{\pi^3 S}}\left(2 \tilde{b}^2 S-2 \sqrt{\tilde{b}^4 S^2+\pi ^2 \tilde{b}^2 q^2}+8 \pi  P S+\pi\right).
\end{eqnarray}
Parameter $\tilde{b}$ represents the maximal electromagnetic field strength, which can also be related to string tension. The corresponding thermodynamic function reads
\begin{eqnarray}
 \Phi=\frac{1}{2\sqrt{\pi S}\sin\theta}\left(1-\frac{2\pi \tilde{b} q^2}{\sqrt{\tilde{b}^2 S^2+\pi ^2 q^2}}\right).
\end{eqnarray}
Then one can easily obtain the vector field $\phi$ and the normalized vector field $n$. We exhibit the behavior of the normalized vector field $n$ in Fig. \ref{pBIVec} with $q$=1 and $\tilde{b}$=0.4 for the charged BI-AdS black hole. For this case, two critical points CP$_2$ and CP$_3$ are found. We construct three contours $C_3$, $C_4$, and $C_5$, which share the same parametrized form as (\ref{pfs}), but with $(a, b, r_0)$=(0.3, 0.3, 1.97), (0.3, 0.3, 3.69), and (1.5, 0.9, 2.83), respectively. We calculate the deflection angle $\Omega(\vartheta)$ for these three contours and the results are shown in Fig. \ref{pBICh}. They present different behaviors. The function $\Omega(\vartheta)$ increases along $C_3$ and decreases along $C_4$, while first decreases, then increases, and finally decreases along $C_5$. $\Omega(2\pi)$=2$\pi$, $-2\pi$, and 0 for these three contours. Therefore, the topological charge $Q_{\text{CP}_2}=1$ and $Q_{\text{CP}_3}=-1$ for the critical points CP$_2$ and CP$_3$. Since they have different values, these two critical points belong to different topological classes. CP$_3$ is the conventional critical point, while CP$_2$ is a novel one. Significantly, the total topological charge for the charged BI-AdS black hole is
\begin{eqnarray}
 Q=Q_{\text{CP}_2}+Q_{\text{CP}_3}=0,
\end{eqnarray}
which equals that along contour $C_5$. Thus the charged black hole system and the BI-AdS black hole system have different topology from the viewpoint of the thermodynamics.

\section{Feature of critical points}

As shown above, there are two types of critical points from the topology. The conventional critical point has $Q$=-1 while the novel one has $Q$=1. One wonders whether there exist difference between them. Here we aim to disclose it.

\begin{figure}
\center{\subfigure[]{\label{BItsCon}
\includegraphics[width=4.1cm]{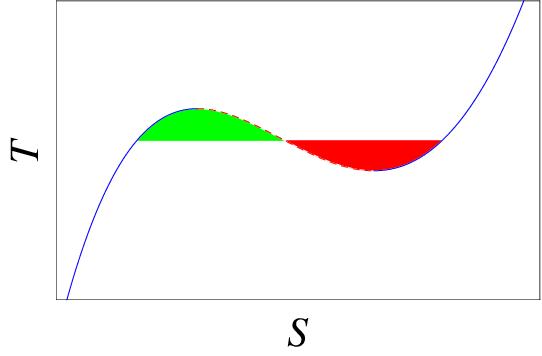}}
\subfigure[]{\label{BItsNovel}
\includegraphics[width=4.1cm]{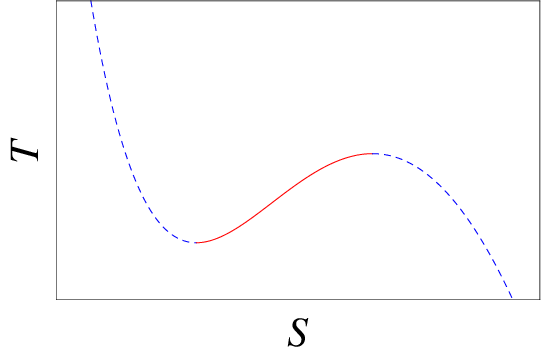}}}
\caption{Behaviors of the isobaric curves for the BI-AdS black hole with $a$=1 and $\tilde{b}$=0.4. Solid and dashed curves are for the thermodynamic stable and unstable black hole branches, respectively. (a) Near the conventional critical point CP$_3$. The shadowed regions have the same area. (b) Near the novel critical point CP$_2$.}\label{aaBItsNovel}
\end{figure}

We show the isobaric curves in Fig. \ref{aaBItsNovel} near the critical points CP$_2$ and CP$_3$. The local thermodynamic stability of a system is determined by the heat capacity $C_P=T(\partial_ST)^{-1}_{P,x^{i}}$. Therefore, the black hole branch with positive slope is stable, while the black hole branch with negative slope is unstable. In the figure, the solid (dashed) curve denotes the stable (unstable) black hole branch.

Near the conventional critical point CP$_3$, we can see that the intermediate black hole branch is unstable, and it can be eliminated by using Maxwell's equal area law. By constructing these two equal areas (in green and red colors), see Fig. \ref{BItsCon}, one can determine the first-order phase transition among these two black hole branches marked with blue solid color. This behavior is well known among the small-large black hole phase transition. In the $P$-$T$ phase diagram, one can observe that the coexistence curve of the first-order phase transition extends from such critical point.

Near the novel critical point CP$_2$, we observe a different pattern shown in Fig. \ref{BItsNovel}. The intermediate black hole branch is stable, while others are unstable. So the Maxwell's equal area law is not applicable for this case, and thus no first-order phase transition can take place.

Therefore, we can conclude that the first-order phase transition can emerge from the conventional critical point with $Q=-1$, while it cannot from the novel critical point with $Q=1$.

\section{Summary}

In this work, we introduced the topology to the study of the critical point in black hole thermodynamics, especially by using the black hole temperature and entropy. Although critical points have been well understood in the study of the phase transition, we emphasized that there are two different types of critical points, which are neglected before. We named them the conventional and novel critical points, respectively, endowed with the topological charge $Q=-1$ and 1.

The different topological charges reveal that these two types of critical points have different topological properties. As an example, we showed that the coexistence curve of the first-order phase transition can only extend from the conventional critical points, while the presence of the novel critical points cannot seve as an indicator of the first-order phase transition.

Moreover, the topological charge for the black hole systems is the sum of the winding number of all its critical points. For the charged AdS black hole with $q=1$, its topological charge $Q=-1$, while for the charged BI-AdS black hole with $q=1$ and $b=0.4$, the topological charge vanishes. This suggests that they belong to different classes of topology from the viewpoint of thermodynamics. Since there are more richer phase structures for some other black hole systems, it is worth generalizing this topological study and we believe more interesting topological properties shall be uncovered. Note that this approach is also applicable for other ordinary thermodynamic systems.

\section*{Acknowledgements}
This work was supported by the National Natural Science Foundation of China (Grants No. 12075103, No. 11675064, No. 11875151, and No. 12047501), the 111 Project (Grant No. B20063).

\section*{Appendix: Critical points}

The critical point can be determined by one of the conditions shown in Eqs. (\ref{cococ})-(\ref{yots}). However, in this paper, we employ a different way. Here we would like to give a brief clarification to show that they are equivalent. For convenience, we take the four-dimensional charged AdS black hole as an example. The generalization will be natural then.

\begin{figure}
\includegraphics[width=7cm]{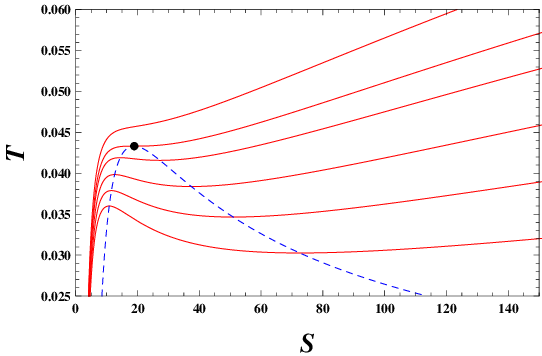}
\caption{Isobaric curves (red solid curves) for the charged AdS black hole shown in the $T$-$S$ plane with $q=1$. The blue dashed curve is for the extremal points of the temperature and the critical point is marked with the black dot.}\label{pIsco}
\end{figure}

By making use of the equation of state (\ref{iscop}), we show the isobaric curves described by the red solid curves in Fig. \ref{pIsco} with $q=1$. As expected, there are two extremal points of the temperature on each isobaric curve with a pressure below its critical value. With the increase of the pressure, these two extremal points coincide at the critical point.

Adopting the condition (\ref{cots}), we easily obtain the critical point
\begin{eqnarray}
 P=\frac{1}{96\pi q^2},\quad T=\frac{\sqrt{6}}{18\pi q},\quad
 S=6\pi q^2,
\end{eqnarray}
which is exactly the result given in Ref. \cite{Kubiznak}, and is marked with the black dot in the figure.

Our approach in this paper is calculating the extremal points of the temperature first. Solving $(\partial_{S}T)_{P,q}=0$, one easily has the pressure
\begin{eqnarray}
 P=\frac{S-3\pi}{8S^2}.
\end{eqnarray}
Bring back into (\ref{iscop}), we obtain the temperature of these extremal points
\begin{eqnarray}
 T=\frac{S-2\pi q^2}{2\sqrt{\pi S^3}},\label{aad}
\end{eqnarray}
which is exactly the function of $\Phi*\sin\theta$, and is described by the blue dashed curve. Further taking $(\partial_{S}T)_q=0$ or $(\partial_{S}\Phi)_q=0$, the critical point will be obtained. Alternatively, we can also find this result from the figure that the critical point obtained by (\ref{cots}) exactly meets the extremal point of the temperature (\ref{aad}) described by the blue dashed curve.

Now we can see that these two methods determining the critical point are equivalent for the small-large black hole phase transition.

\end{document}